# An approach to find dynamic slice for C++ Program

## Santosh Kumar Pani and Priya Arundhati

**Abstract**—Object-oriented programming has been considered a most promising method in program development and maintenance. An important feature of object-oriented programs (OOPs) is their reusability which can be achieved through the inheritance of classes or reusable components.Dynamic program slicing is an effective technique for narrowing the errors to the relevant parts of a program when debugging. Given a slicing criterion, the dynamic slice contains only those statements that actually affect the variables in the slicing criterion. This paper proposes a method to dynamically slice object-oriented (00) programs based on dependence analysis. It uses the Control Dependency Graph for object program and other static information to reduce the information to be traced during program execution. In this paper we present a method to find the dynamic Slice of object oriented programs where we are finding the slices for object and in case of function overloading.

——————————— ◆ ———————————

## 1 INTRODUCTION

Program slicing is an effective technique for narrowing the focus of attention to the relevant parts of a program. The slice of a program consists of those statements and predicates that may directly or indirectly affect the variables computed at a given program point [1]. The program point s and the variable set V, denoted by <s, V>, is called a slicing criterion. The basic idea of program slicing is to remove irrelevant statements from source codes while preserving the semantics of the program such that at the specified point the variable produces the same value as its original program. Program slicing has been widely used in many software activities, such as software analyzing, understanding, debugging, testing, maintenance, and so on [2, 3, 4]. Slicing algorithm can be classed according to whether they only use static information (static slicing) or dynamic execution information for a specific program input (dynamic slicing). This paper focuses on the dynamic slicing methods for finding the slices for class,object and in case of function overloading.

## 2 REVIEW OF RELATED WORKS

Agrawal proposes a dynamic slicing method by marking nodes or edges on a static program dependence graph during execution [5]. The result is not precise, because some dependencies might not hold in dynamic execution. Agrawal also proposes a precise method based on the dynamic dependence graph (DDG) [5], and Zhao applies it to slice object-oriented programs [6]. The shortcoming is that the size of the DDG is unbound.

Korel [8], Song [7] and Tibor [9] propose forward dynamic slicing methods and Song also proposes method to slice

—————————————————

- *Santosh Kumar Pani is with the School of Computer Engineering, KIIT University, Bhubaneswar.*
- *Priya Arundhati is with the Department of CSE, Nalanda Institute of Technology, Bhubaneswar.*

OOPs using dynamic object relationship diagram (DORD). In these methods, they compute the dynamic slices for each statement immediately after this statement is executed. After the last statement is executed, the dynamic slices of all statements executed have been obtained. However, only some special statements in loops need to compute dynamic slices. In this paper we represent the object oriented program in control dependence Graph (CDG).

## 3 BASIC CONCEPT & DEFINITION

Slicing object oriented program is quite difficult as it comprises of many features like Class, Object, Polymorphism, inheritance etc. In this paper we are trying to finding the slices for object and in case of function overloading.

In object oriented programming language it is possible to define more then one function that share a common name this feature is called function overloading.Hence our aim is to find the Slice where the function is overloaded.In that case the Slicer should know that which version of the function to call.For this,the Slicer uses the following information about the function Call.

1. The type of the argument.
2. The no.of argument.
3. The name of the function.

After getting these information from the calling node the unquiue function will be called and the slice of the function will be found out in normal method.
The following Data Structures are used for this purpose.

1. Control Dependence Graph: The control dependence graph (CDG) G of an object oriented program P is a graph G=(N,E),where each node $n \in N$ represents a statement of the program P. For any pair of nodes x and y,(x,y)εE if node x is control dependent on node y.

2. Def(dvar) : If dvar is a variable in a program P.A node u of CDG Gp is said to be a Def(dvar) node if u is definition statement that defines the variable dvar.

3. Use(dvar) : If dvar is a variable in a program P.A node u of CDG Gp is said to be a Use(dvar) node if he statement u uses the value of the variable dvar.

4. Defvarset(u):Let u be a node in the CFG Gp of a program P.The set DefVarSet(u)={dvar:dvar is a data variable in the program P,and u is a Def(dvar) node

5. Usevarset(u):Let u be a node in the CFG Gp of program P. The set UseVarSet(u)={dvar:dvar is a data variable in the program P,and u is a Use(dvar) node}.

6. ActiveControlSlice(s): Let s be a test node(predicate statement) in, the CDG Gp of a program P and UseVar-Set(s)={var1,var2,…..vark}. Before execution of the program P, ActiveControlSlice(s) = ∅.After each execution of the node s in the actual run of the program, ActiveControl-Slice(s)={s}UActiveDataSlice(var1)…UActiveDataSlice(vark) U ActiveControlSlice(t), where t is most recently executed successor node of s in Gp.If s is a loop control node and the present execution of the node s corresponds to exit from loop,then ActiveControlSlice(s)= ∅

7. ActiveDataSlice(var):Let dvar be a data variable in the object oriented program P. Before execution of the program P ,ActiveDataSlice(obj.var)=∅.Let u be a Def(obj.var) node, and UseVar-Set(u)={var1,var2,..vark}.After each execution of the node u in the actual run of the program, Ac-tiveDataSlice(obj.var)={s}U ActiveDataSlice(var1) U………UActiveDataSlice(dvark) U ActiveCon-trolSlice(t), where t is most recently executed successor node of s in Gp.

/*For each data member of an object a separate slice will be mainted in a shared memory which is only accessable to the member function called on that object.*/

8. DyanSlice(obj):-Let obj be an object of the class in the proram P. Before execution of program P, DyanSlice(obj)=φLet the object contains the data member mem1.mem2,…memn then Dyan-Slice(obj)=DyanSlice(obj.mem1) U Dyan-Slice(obj.mem2)U…..U DyanSlice(obj.memn)

9. DyanSlice(s,var):-Let s be a node of the CDG of a program P,& var be a variable in the set i.e. var(-DefVarSet(s) U UseVarSet(s) Before executionof the program P DyanSlice(s,var)=φAfter each ex-ecution of the node s in an actual run Dyan-Slice(s,var)=Active Data Slice(var) U Active Con-trol Slice(t)

10. Active Call Slice:-Let Gp represents the CDG of the multiprocedure program P. Before execution of the program P Active Call Slice=φLet Uactive be an active Call node then Active Call Slice={Uactive} U Active Call Slice U Active Con-trol Slice(t).

11. Call Slice Stack:-This stack is maintained to keep track of the Active Call slice during the actual run of the program.

12. Formal(x,var),Actual(x,var)): Let p1 be a proce-dure of a program p having multiple procedures, and x be a calling node to the procedures, and x be a calling node to the procedure p1.Let f be a formal parameter of procedure P1 and its corres-ponding actual parameter at the calling node to the procedure P1 and its corresponding actual parameter at the calling node x be a. We define Formal(x,a)=f and Actual(x,f)=a.

# 4 AN ALGORITHM FOR FINDING DYNAMIC SLICE FOR OBJECT ORIENTED PROGRAMS IN CASE OF FUNCTION OVERLOADING.

**Step 1:** Start

**Step2:** At first Control Dependence Graph G of object oriented program P is statically constructed .

**Step 3:** Before each execution of the program do step 4 to step 7.

**Step 4:** For each node u of G do step 5 to step 6.

**Step 5 :** If u is a test node, then ActiveControlSlice(u) = ∅.

**Step 6:** For each variable var εDefVarSet(u) U UseVar-Set(u) do DyanSlice(u,var) = ∅.

**Step 7:** CallSliceStack=NULL.
       ActiveCallSlice= ∅.

**Step 7a.** For each data variable dvar of the program
       do
         ActiveDataSlice(var)= ∅.

**Step 7b.** If var is a data member of an corresponding class of an object to then
       do
         ActiveDataSlice(obj.var) = ∅
         for each member declared with in the class of an object

**Step 8:** Run the program P with the given set of input values and repeat steps 9 to 19 until the program termi-nates.

**Step 9:** Before execution of each call node u do step 10 and step 11

**Step 10a.** Let u be a call node to a procedure Q, update CallSliceStack and ActiveCallSlice.

//In case of function having same name the signature of the function is foundout as each function have unquiue signature ,hence the slicer will know that which version of the function to call.//

**Step 10b.** For each actual parameter var in the procedure call Q do ActiveDataSlice (Formal(u,var)) = ActiveDataSlice(var) U ActiveCallSlice.

//Here when a member finction of an class is called with an object then the actual parameter of the function is individually copied to the formal parameter of the member function.//

**Step 11:**Update ActiveReturnSlice before execution of the return node.

**Step 12:** After execution of each node u of the program P do step 13 to step 18.

**Step 13:**If u is a Def(var) node and not a call node then Update ActiveDataSlice(var).

//var can be a data variable or data member of an object then Update ActiveDataSlice(var) will be done accordingly //

**Step 14:** If u is a call node to a procedure Q then for every formal reference parameter var in the procedure Q do ActiveDataSlice(Actual(u,var))= ActiveDataSlice(var).

**Step 15:** If u is a Def(var) node then ActiveDataSlice(var)=ActiveReturnSlice.

**Step16a.** For every variable var if declared as automatic local in the procedure Q do ActiveDataSlice= Ø.

//For local static variable and global variable the ActiveDataSlice remains unchanged

**Step16b.** Update CallSliceStack and ActiveCallSlice and Set ActiveReturnSlice= Ø .

**Step 17:** For every variable var ε DefVarSet(u) U UseVarSet(u) Update DyanSlice(u,var).

**Step 18:** If u is a test node, then update ActivecontrolSlice(u).

**Step 19:**Exit

## 5 SAMPLE PROGRAM AND WORKING OF THE ALGORITHM

Sample Program

The following object oriented Program where only the concept of class and object are used is discussed to iilustrate our proposed Algorithm.

```
Class  test
{
int a;
int b;
public:
void get(int x,int y)
{
17.a=x;
18.b=y;
}
Void display()
{
19.Cout<<a;
20.Cout<<b;
}
Test add (test tp1,test tp2)
{
21.a=tp1.a+tp2.a;
22.b=tp1.b+tp2.b;
}
test  add(test tp3,int s)
{
23.a=tp3.a +s;
24.b=tp3.b +s;
}
};

void main()
{
test T1,T2,T3,T4;
int p,q;
1.Cout<<"Enter the value of p";
2.Cin>>p;
3.Cout<<"Enter the value of q";
4.Cin>>q;
5.T1.get(p,q);
6.T1.display();
7.Cout<<"Enter the value of p";
8.Cin>>p;
9.Cout<<"Enter the value of q";
10.Cin>>q;
11.T2.get(p,q)
12.T2.display()
13.T3.add(t1,t2);
14.T3.display();
15.T4.add(T3,5);
16.T4.display();
}
```

Control Dependency Graph

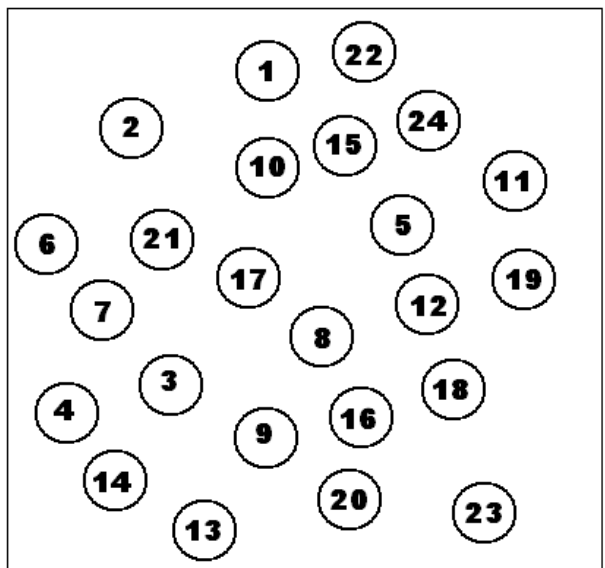

**Control Dependency graph of the sample program**

**Working of the Algorithm**

**After Execution of node-2**
ADS(p)={2}
DyanSlice(2,p)={2}
**After Execution of node-4**
ADS(q)={4}

DyanSlice(4,q)={4}

**After Execution of node-5**

Here one of the object of test class is T1.get() is a member function of the test class.The formal parameter of get() is x & y as well as these are the local variable to get().The actual parameter is p & q.

So,ADS(x)=ADS(p) U Active Call Slice
            ={2} U {5}={2,5}

ADS(y)=ADS(q) U Active Call Slice
            ={4} U{5}={4,5}

Again T1 is calling the get() so here a & b are T1's variable

ADS(T1.a)=ADS(x)U{15}={2,5} U{15}={2,5,17}

DyanSlice(15,T1.a)={2,5,17}

ADS(T1.b)=ADS(y)U{16}={4,5,18}

DyanSlice(16,T1.b)={4,5,18}

**After Execution of node-6**

Now T1 is calling display()

So,DyanSlice(17,T1.a)=ADS(T1.a)={2,5,17}

DyanSlice(18,T1.b)=ADS(T1.b)={4,5,18}

DyanSlice(T1)=DyanSlice(T1.a) U DyanSlice(T1.b)
            ={2,5,17} U {4,5,18}
            ={2,4,5,17,18}

**After Execution of node-8**

ADS(p)={8}

DyanSlice(8,p)={8}

**After Execution of node-10**

ADS(q)={10}

DyanSlice(10,q)={10}

**After Execution of node-11**

ADS(x)=ADS(p) U Active Call Slice
            ={8} U{11}={8,11}

ADS(y)=ADS(q) U Active Call Slice
            ={10}U{11}={10,11}

ADS(T2.a)=ADS(x) U 17
            ={8,11,17}

DyanSlice(17,T2.a)={8,11,17}

ADS(T2.b)=ADS(y) U 18
            ={10,11} U 18={10,11,18}

**After Execution of node-12**

DyanSlice(19,T2.a)=ADS(T2.a)={8,11,17}

DyanSlice(20,T2.b)=ADS(T2.b)={10,11,18}

DyanSlice(T2)=DyanSlice(T2.a) U DyanSlice(T2.b)
            ={8,11,17} U {10,11,118}
            ={8,10,11,17,18}

**After Execution of node-13**

Here T3 is calling the add() where the formal parameter are tp1 & tp2 but here the object T1 & T2 are passed as the actual parameter.and there are two version of add()function.Now the Slicer will decide which function to call by checking the signature of the function.when this node will be executed the control will go to the first add()function where the argument are the objects.

So,ADS(tp1.a)=ADS(T1.a) U Active Call Slice
            ={2,,5,17} U {13}
            ={2,5,17,13}

ADS(tp1.b)= ADS(T1.b) U Active Call Slice
            ={4,5,18} U 13
            ={4,5,18,13}

ADS(tp1)=ADS(tp1.a)  U  ADS(tp1.b)={  2,5,17,13}      U
{4,5,18,13}={ 2,5,17,13,4,5,18,13}

ADS(tp2.a)=ADS(T2.a) U Active Call Slice
            ={8,11,17} U {13}
            ={8,11,17,13}

ADS(tp2.b)= ADS(T2.b) U Active Call Slice
            ={10,11,18} U 13
            ={10,11,18,13}

ADS(tp2)=ADS(tp2.a)  U  ADS(tp2.b)={  8,11,17,13}      U
{10,11,18,13}={ 8,10,11,13,17,18}

ADS(T3.a)=ADS(tp1.a) U ADS(tp2.a)  U {21}
            ={2,,5,17} U {8,11,17,13} U{21}
            ={2,5,8,11,13,17,21}

ADS(T3.b)= ADS(tp1.b) U ADS(tp2.b) U{20}
            ={4,5,13,18} U {10,11,13,18} U{22}
            ={4,5,10,11,13,18,22,}

DyanSlice(13,T3.a)=ADS(T3.a)={2,5,8,11,13,17,21,}

DyanSlice(13,T3.b)=ADS(T3.b)={4,5,10,11,13,18,22}

**After Execution of node-14**

DyanSlice(19,T3.a)=ADS(T3.a)={ 2,5,8,11,13,17,21}

DyanSlice(18,T3.b)=ADS(T3.b)={ 4,5,10,11,13,18,22}

DyanSlice(T3)=DS(T3.a) U DS(T3.b)
            =            {2,5,8,11,13,17,21}      U
{4,5,10,11,13,18,22}={2,4,5,8,10,11,13,17,18,21,22}

**After Execution of node-15**

ADS(tp3.a)=ADS(T3.a) U Active Call Slice
            ={2,5,8,11,13,17,21} U 15
            ={2,5,8,11,13,15,17,21}

ADS(tp3.b)=ADS(T3.b) U Active Call Slice
            ={4,5,10,11,13,18,22,} U15
            ={4,5,10,11,13,15,18,22,}

ADS(T4.a)=ADS(tp3.a) U Active Call Slice
            ={2,5,8,11,13,15,17,21} U{23}
            ={2,5,8,11,13,15,17,21,23}

ADS(T4.b)=ADS(tp3.b) U Active Call Slice
            ={4,5,10,11,13,15,18,22,} U {24}
            ={4,5,10,11,13,15,18,22,24}

**After Execution of node-16**

DyanSlice(16,T4.a)=ADS(T4.a)
                  ={2,5,8,11,13,15,17,21,23}

DyanSlice(16,T4.b)=ADS(T4.b)
                  ={4,5,10,11,13,15,18,22,24}

DyanSlice(T4)= DyanSlice(T4.a) U  DyanSlice(T4.b)
            ={2,5,8,11,13,15,17,21,23}          U
{4,5,10,11,13,15,18,22,24}
            ={2,4,5,8,10,11,13,17,18,21,22,15,23,24}

# 6   ANALYSIS OF THE ALGORITHM

The improved InterSlice Algorithm  also uses a collection of control dependence graphs as the intermediate program representation & computes precise dynamic slice. The space complexity of the improved Interslice algorithm is also O(n2)The time complexity of this algorithm is linear as no extra data structures are required  for this enhancement .This algorithm does not require to store the past execution histo-ry.

## 7 CONCLUSION

In this paper of the dynamic Slicing of Object Oriented Programs is discussed. The Object Oriented Programs has various features like Class, Object, Dynamic Binding, Polymorphism, Inheritance etc. In this proposed Algorithm only the Class, Object and function overloading are discussed. Finding the Slices for Inheritance, Dynamic binding and other features are the future work of this Research